\documentclass[11pt]{article}
\usepackage[english]{babel}
\usepackage[utf8]{inputenc}
\usepackage[T1]{fontenc}
\usepackage{amsmath, amssymb, amsthm}
\usepackage{graphicx}
\usepackage{hyperref}
\hypersetup{
    colorlinks=true,
    linkcolor=blue,
    filecolor=magenta,
    urlcolor=cyan,
}
\usepackage{listings}
\usepackage{xcolor}
\definecolor{codegreen}{rgb}{0,0.6,0}
\definecolor{codegray}{rgb}{0.5,0.5,0.5}
\definecolor{codepurple}{rgb}{0.58,0,0.82}
\definecolor{backcolour}{rgb}{0.95,0.95,0.92}
\lstdefinestyle{mystyle}{
    backgroundcolor=\color{backcolour},
    commentstyle=\color{codegreen},
    keywordstyle=\color{magenta},
    numberstyle=\tiny\color{codegray},
    stringstyle=\color{codepurple},
    basicstyle=\ttfamily\footnotesize,
    breakatwhitespace=false,
    breaklines=true,
    captionpos=b,
    keepspaces=true,
    numbers=left,
    numbersep=5pt,
    showspaces=false,
    showstringspaces=false,
    showtabs=false,
    tabsize=2
}
\title{A Semantic Model for Audit of Cloud Engines based on ISO/IEC TR 3445:2022}
\author{Morteza Sargolzaei Javan \\ Amirkabir University of Technology \\ \texttt{msjavan@aut.ac.ir}}
\date{2025-10-09}

\begin{document}
\maketitle
\begin{abstract}
Cloud computing has become the foundation of modern digital infrastructure, yet the absence of a unified architectural and compliance framework impedes interoperability, auditability, and robust security. This paper introduces a formal, machine-readable semantic model for Cloud Engines, integrating the architectural taxonomy of ISO/IEC 22123 (Cloud Reference Architecture) with the security and compliance controls of ISO/IEC 27001:2022 and ISO/IEC TR 3445:2022. The model decomposes cloud systems into four canonical interfaces—Control, Business, Audit, and Data—and extends them with a security ontology that maps mechanisms such as authentication, authorization, and encryption to specific compliance controls. Expressed in RDF/Turtle, the model enables semantic reasoning, automated compliance validation, and vendor-neutral architecture design. We demonstrate its practical utility through OpenStack and AWS case studies, and provide reproducible validation workflows using SPARQL and SHACL. This work advances the state of cloud security modeling by bridging architectural and compliance standards in a unified framework, with a particular emphasis on auditability.
\end{abstract}

\section{Introduction}

Cloud computing underpins a vast array of digital services, yet the diversity of interfaces and the complexity of compliance requirements present persistent challenges for system architects and operators. Existing standards typically address either functional APIs (e.g., OCCI for resource control) or security policies (e.g., ISO/IEC 27001) in isolation, resulting in fragmented approaches to architecture and compliance.

To address this gap, we propose a holistic semantic model for \textit{Cloud Engines} that unifies architectural and security perspectives. The model is grounded in ISO/IEC 22123 \cite{iso22123}, which provides a vendor-neutral reference architecture and service taxonomy, and is extended with a security ontology that maps mechanisms to controls in ISO/IEC 27001 \cite{iso27001}, NIST SP 800-53 \cite{nist80053}, CSA CCM \cite{csa-ccm}, and cloud provider frameworks such as the AWS Well-Architected Framework \cite{aws-waf}. We decompose cloud systems into four canonical interface categories:

\begin{itemize}
  \item \textbf{Control Interface}: Lifecycle management of resources (for example, OCCI).
  \item \textbf{Business Interface}: User-facing operations such as billing, dashboards, and Single Sign-On (SSO).
  \item \textbf{Audit Interface}: Emission of logs and metrics for monitoring and compliance (e.g., syslog, CloudTrail, StatsD).
  \item \textbf{Data Interface}: Persistent data storage and access (e.g., S3, Swift, NFS).
\end{itemize}

A comprehensive cloud architecture specification must not only define these interfaces, but also prescribe their security properties and compliance mappings. Our principal contribution is an RDF-based ontology that formally defines these interfaces, their security attributes, and their alignment with established industry standards.

SmartData 4.0 \cite{Sargolzaei2019} provides a framework to describe big data problems and solutions in a formal language, accelerating innovation and development across various sectors. This framework enables the formalized description of data operations such as data fusion, transformation, and provenance management, empowering raw data with intelligence.

\paragraph{From model to validation}
The workflow presented in this paper comprises: (1) authoring a semantic model that specifies required interfaces and policies; (2) instantiating the model with concrete services and configurations (e.g., mapping OpenStack Keystone, Swift, and Ceilometer to the model); (3) executing automated compliance checks using SPARQL queries and SHACL validations to identify gaps; and (4) generating actionable reports or remediation tasks for operations teams. Worked examples and validation snippets are provided in the Appendix to facilitate reproducibility.

This paper aims to integrate the concepts of SmartData 4.0 with the CloudEngine framework to establish a standard and intelligent cloud engine. By leveraging the formal description capabilities of SmartData 4.0, we can enhance the CloudEngine framework to move towards the realization of intelligent and autonomous clouds. This integration not only aligns with the principles of ISO/IEC 22123 but also paves the way for innovative cloud solutions that are context-aware and self-governing.

\section{Background and Related Work}

\subsection{Cloud Interface Standards}
The Open Cloud Computing Interface (OCCI) is a RESTful protocol and API standard developed by the Open Grid Forum to manage cloud infrastructure resources. OCCI addresses the control plane but does not by itself prescribe a complete architecture encompassing business, audit, and data planes.

While OCCI provides a robust foundation for control plane operations, it does not encompass the full spectrum of architectural requirements, such as business, audit, and data interfaces. This limitation underscores the need for a more comprehensive model that integrates functional and security perspectives.

\subsection{Security and Compliance Frameworks}
Modern cloud security is governed by multiple overlapping standards. Representative frameworks include:
\begin{itemize}
  \item \textbf{ISO/IEC 27001:2022}: A framework for an Information Security Management System (ISMS) with a comprehensive control set \cite{iso27001}.
  \item \textbf{NIST SP 800-53 Rev. 5}: A catalog of security and privacy controls organized into families such as Access Control (AC), System and Communications Protection (SC), Audit and Accountability (AU), etc. \cite{nist80053}.
  \item \textbf{Cloud Security Alliance (CSA) CCM v4}: A cloud-focused control matrix designed to harmonize with other standards like ISO and NIST \cite{csa-ccm}.
  \item Vendor-specific guidance, e.g., the \textbf{AWS Well-Architected Framework} \cite{aws-waf}.
\end{itemize}

Our work synthesizes these frameworks into a unified, standards-aligned model that is applicable across diverse cloud environments and deployment scenarios.

Cloud security and compliance are governed by a constellation of overlapping standards and frameworks, each addressing different aspects of risk management, control implementation, and auditability. Key frameworks include:
\begin{itemize}
  \item \textbf{ISO/IEC 27001:2022}: Establishes requirements for an Information Security Management System (ISMS) and provides a comprehensive set of controls for organizational security.
  \item \textbf{NIST SP 800-53 Rev. 5}: Offers a detailed catalog of security and privacy controls, organized into families such as Access Control (AC), System and Communications Protection (SC), and Audit and Accountability (AU).
  \item \textbf{Cloud Security Alliance (CSA) CCM v4}: Presents a cloud-specific control matrix designed to harmonize with ISO, NIST, and other standards, facilitating cross-framework compliance.
  \item \textbf{AWS Well-Architected Framework}: Provides vendor-specific best practices for secure cloud architecture and operations.
\end{itemize}

Our model synthesizes these frameworks into a unified ontology, enabling explicit mapping of architectural components and security mechanisms to compliance controls across heterogeneous cloud environments.

\paragraph{On ISO/IEC 22123}
ISO/IEC 22123 (Cloud Reference Architecture) provides a complementary architectural viewpoint focused on service taxonomy, functional blocks, and interaction patterns for cloud systems. While ISO/IEC 27001 prescribes the controls and management processes for an organization's information security management system (ISMS), ISO/IEC 22123 helps architects map those controls to concrete cloud functions and interfaces. In this paper we leverage ISO/IEC 22123 to ground our interface taxonomy (Control, Business, Audit, Data) in a vendor-neutral cloud reference architecture, and then map the resulting components to security controls drawn from ISO/IEC 27001, NIST, CSA, and cloud provider guidance.

\section{The Cloud Engine Model}

The model is expressed in RDF/Turtle. We define a namespace \texttt{cloudeng:} for core cloud concepts and \texttt{sec:} for security concepts. Industry standards are referenced via conceptual namespaces (e.g., \texttt{iso27001:}, \texttt{nist80053:}).

\subsection{Core Architecture}
The foundational class is \texttt{cloudeng:CloudEngine}, which aggregates four interface types. The model defines object properties such as \texttt{cloudeng:hasControlInterface} to link an engine to its interfaces.

\subsection{Security Ontology}
We extend the core model with a security layer that includes classes for identity providers, authentication mechanisms, authorization mechanisms, encryption methods, and transport security. Typical classes include:
\begin{itemize}
  \item \texttt{sec:IdentityProvider} (e.g., Keycloak, Okta)
  \item \texttt{sec:AuthenticationMechanism} (e.g., OAuth 2.0, SAML)
  \item \texttt{sec:AuthorizationMechanism} (e.g., RBAC, ABAC)
  \item \texttt{sec:EncryptionMethod} (e.g., AES-256, TLS 1.3)
\end{itemize}

Each interface instance can be annotated with these security properties using RDF properties such as \texttt{sec:supportsAuthentication} and \texttt{sec:encryptsData}.

\subsection{Standards Alignment}
A key property is \texttt{sec:implementsStandard}, which allows any security mechanism or cloud service to be explicitly linked to the specific controls it satisfies in target standards. For example, RBAC can be linked to ISO 27001 control A.9.4.1 (information access restriction) and NIST control AC-3 (access enforcement). A high-level \texttt{sec:SecurityPolicy} can then declare compliance with a set of standards using \texttt{sec:compliesWith}.

A central feature of our ontology is the \texttt{sec:implementsStandard} property, which enables explicit linkage between security mechanisms, cloud services, and the compliance controls they satisfy in target standards. For example, Role-Based Access Control (RBAC) can be mapped to ISO/IEC 27001 control A.9.4.1 (information access restriction) and NIST SP 800-53 control AC-3 (access enforcement). High-level \texttt{sec:SecurityPolicy} instances can declare compliance with multiple standards using \texttt{sec:compliesWith}, supporting multi-framework validation and reporting.

\begin{itemize}
  \item \textbf{ISO/IEC 22123 (Cloud Reference Architecture)}: Provides a functional decomposition and service taxonomy for cloud systems. We leverage ISO/IEC 22123 to map our four-interface model to canonical cloud components (e.g., identity and control planes, data plane, and telemetry/audit plane), facilitating precise placement of ISO/IEC 27001 controls within operational architectures.
\end{itemize}

\section{Model Implementation and Examples}
We instantiate the model with real-world examples to demonstrate practical utility, including OpenStack components (Keystone, Swift, Ceilometer) and AWS services (IAM, S3, CloudTrail). These instances show how concrete services map to interface types and to security controls.

To demonstrate the practical utility of our semantic model, we instantiate it with real-world examples from OpenStack and AWS. These case studies illustrate how concrete services map to interface types and compliance controls, and how the ontology supports automated validation and reporting.

\subsection{OpenStack mapping}
To make the model actionable for OpenStack operators, we provide a concise mapping and operational notes for common components:

\begin{itemize}
  \item \textbf{Keystone (Identity / Control)}: Keystone provides authentication (tokens, federation), identity management (users, groups, domains), and role assignments. In the model Keystone instances map to both \texttt{cloudeng:ControlInterface} (APIs for creating projects/users/roles) and \texttt{sec:IdentityProvider}. Important operational attributes to capture in the instance are Keystone API version (v3), token backend (Fernet vs PKI), federation configuration (mapped IdPs), and whether application credentials or trust relationships are enabled.

  \item \textbf{Policy and Authorization}: OpenStack services rely on policy files (\texttt{policy.json} or \texttt{policy.yaml}) that express access rules. The model should represent service-specific policy rules (e.g., as \texttt{sec:PolicyRule} or linking to a \texttt{cloudeng:policyFile}) so that compliance checks can verify critical rules (such as preventing cross-tenant administrative operations) are present.

  \item \textbf{Swift / Object Storage (Data)}: Swift (or an S3-compatible gateway) maps to \texttt{cloudeng:DataInterface}. Key properties include encryption-at-rest (SSE), key management integration (Barbican or external KMS), object versioning, and public/private container policies. The model should link data interfaces to key-management entities via \texttt{sec:usesKMS} and record whether keys are HSM-backed.

  \item \textbf{Ceilometer / Telemetry (Audit)}: Telemetry pipelines (Ceilometer/Gnocchi/Aodh) and logging (rsyslog/journald → fluentd → central store) should be modeled at two levels: event capture points and long-term storage/retention policies. For compliance we recommend modeling retention duration, integrity controls (append-only or signed logs), and centralized aggregation endpoints.

  \item \textbf{Neutron (Network Isolation)}: Network segmentation and security groups are core to tenancy isolation. The model should capture whether isolated tenant networks, provider networks, or microsegmentation solutions are used, for verifying network-level controls.

  \item \textbf{Barbican / KMS (Key lifecycle)}: Key creation, rotation policy, and custody (HSM-backed or software) influence compliance. Represent KMS as a \texttt{sec:KeyManagement} entity with properties such as rotation frequency and hardware-backed status.
\end{itemize}

\paragraph{Extracting facts from OpenStack}
We recommend a pragmatic approach to instantiate the model from a live OpenStack environment:
\begin{enumerate}
  \item Use the \texttt{openstack} CLI or \texttt{openstacksdk} to export users, projects, role assignments, endpoints, and service configurations as JSON.
  \item Convert the JSON to RDF triples (Turtle) using a small transformation script (example: Python + rdflib). Include triples for service version, policy file contents (or hashes), and KMS linkage.
  \item Load the resulting Turtle file into a triple store and run SHACL validations and SPARQL queries as described in the Appendix.
\end{enumerate}

\paragraph{Operational caveats}
OpenStack deployments vary in topology and version; therefore model instantiations must be version-aware. Where possible include \texttt{cloudeng:serviceVersion} or similar metadata so that compliance checks can account for behavior differences between versions (for example, Keystone v2 vs v3 differences in scoping and federation).

\section{Discussion and Future Work}
The model enables several advanced use cases:
\begin{itemize}
  \item \textbf{Automated Compliance Checking}: SPARQL queries can verify whether a proposed architecture satisfies controls required by a standard (e.g., GDPR Article 32).
  \item \textbf{Architecture Validation}: SHACL shapes can enforce model constraints, such as requiring encryption for all DataInterfaces.
  \item \textbf{Vendor-Agnostic Design}: Architects can design at the semantic level and select implementations that meet the mapped controls.
\end{itemize}

Future work includes adding threat-model constructs (e.g., MITRE ATTACK integration), tooling to generate SHACL from policy definitions, and publishing the ontology with a stable URI for community adoption.

\section{Conclusion}
We presented a standards-aligned semantic model for secure Cloud Engines. By formalizing interfaces and security mappings and expressing them in RDF/Turtle, the model supports reasoning, automated compliance checks, and vendor-agnostic architecture design.

\section{Ontology Overview}
\begin{figure}[h!]
    \centering
    \includegraphics[width=0.8\textwidth]{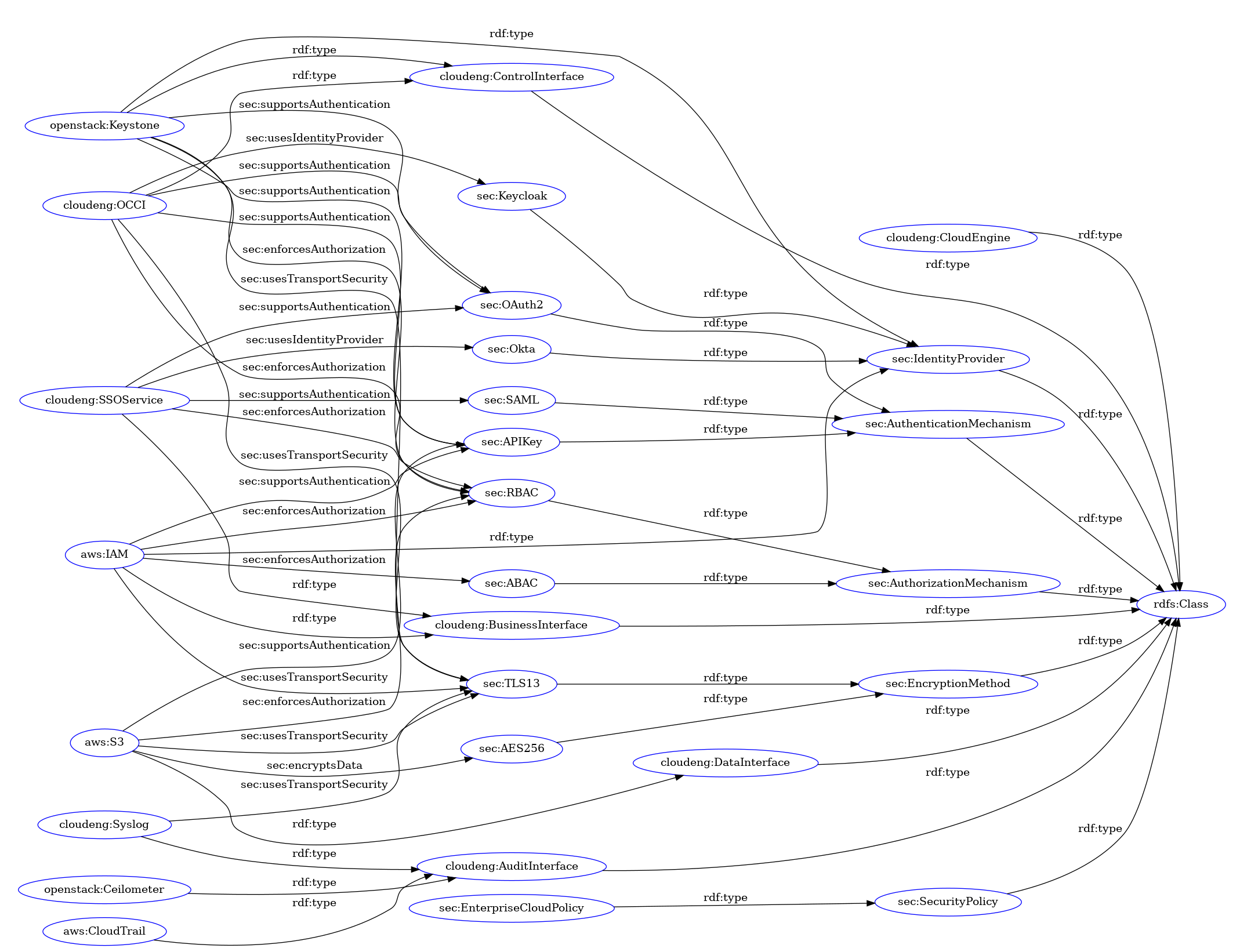}
    \caption{High-level Ontology Graph for CloudEngine}
    \label{fig:ontology-graph}
\end{figure}

To provide a high-level understanding of the ontology, Figure~\ref{fig:ontology-graph} illustrates the core components and their relationships within the CloudEngine framework.

\section{Ontology Implementation}
To facilitate ontology editing and exploration, we provide a Protege-compatible Turtle file that captures the semantic model described in this paper. The file can be imported into Protege for further refinement and validation. The Turtle file is available as \texttt{cloudengine\_protege.ttl} in the supplementary materials.

\appendix
\section{Full RDF/Turtle Model}
\begin{lstlisting}[language={}, caption={Full Cloud Engine Model in RDF/Turtle}]
@prefix rdf:     <http://www.w3.org/1999/02/22-rdf-syntax-ns#> .
@prefix rdfs:    <http://www.w3.org/2000/01/rdf-schema#> .
@prefix xsd:     <http://www.w3.org/2001/XMLSchema#> .
@prefix cloudeng: <http://example.org/cloudengine#> .
@prefix sec:     <http://example.org/security#> .

% Industry Standard Conceptual Namespaces
@prefix iso27001: <https://www.iso.org/standard/27001#> .
@prefix nist80053: <https://csrc.nist.gov/publications/detail/sp/800-53/rev-5/final#> .
@prefix aws:     <https://aws.amazon.com/architecture/well-architected#> .
@prefix openstack: <https://docs.openstack.org/#> .
@prefix gdpr:    <https://eur-lex.europa.eu/legal-content/EN/TXT/?uri=CELEX:32016R0679#> .
@prefix csa:     <https://cloudsecurityalliance.org/artifacts/cloud-controls-matrix/#> .

% ==============================
% CORE CLOUD ENGINE CLASSES
% ==============================

cloudeng:CloudEngine
  a rdfs:Class ;
  rdfs:label "Cloud Engine" ;
  rdfs:comment "A system that provides cloud infrastructure and services." .

cloudeng:Interface
  a rdfs:Class ;
  rdfs:label "Interface" ;
  rdfs:comment "A generic interface through which the cloud engine interacts with external systems." .

cloudeng:ControlInterface
  a rdfs:Class ;
  rdfs:subClassOf cloudeng:Interface ;
  rdfs:label "Control Interface" ;
  rdfs:comment "Interface for managing cloud resources (e.g., provisioning, orchestration)." .

cloudeng:BusinessInterface
  a rdfs:Class ;
  rdfs:subClassOf cloudeng:Interface ;
  rdfs:label "Business Interface" ;
  rdfs:comment "Interface for business operations like billing, SSO, user dashboards." .

cloudeng:AuditInterface
  a rdfs:Class ;
  rdfs:subClassOf cloudeng:Interface ;
  rdfs:label "Audit Interface" ;
  rdfs:comment "Interface for logging, monitoring, and compliance reporting." .

cloudeng:DataInterface
  a rdfs:Class ;
  rdfs:subClassOf cloudeng:Interface ;
  rdfs:label "Data Interface" ;
  rdfs:comment "Interface for data access and storage protocols." .

% ==============================
% SECURITY CLASSES
% ==============================

sec:SecurityPolicy
  a rdfs:Class ;
  rdfs:label "Security Policy" ;
  rdfs:comment "A set of rules and practices that govern security behavior." .

sec:IdentityProvider
  a rdfs:Class ;
  rdfs:label "Identity Provider" ;
  rdfs:comment "Entity that creates, maintains, and manages identity information." .

sec:AuthenticationMechanism
  a rdfs:Class ;
  rdfs:label "Authentication Mechanism" ;
  rdfs:comment "Method used to verify identity (e.g., OAuth2, SAML, API keys)." .

sec:AuthorizationMechanism
  a rdfs:Class ;
  rdfs:label "Authorization Mechanism" ;
  rdfs:comment "Method used to enforce access control (e.g., RBAC, ABAC)." .

sec:EncryptionMethod
  a rdfs:Class ;
  rdfs:label "Encryption Method" ;
  rdfs:comment "Algorithm or standard used for encryption." .

sec:EncryptionScope
  a rdfs:Class ;
  rdfs:label "Encryption Scope" ;
  rdfs:comment "Where encryption is applied (e.g., at-rest, in-transit)." .

sec:TransportSecurityProtocol
  a rdfs:Class ;
  rdfs:label "Transport Security Protocol" ;
  rdfs:comment "Protocol securing data in transit (e.g., TLS, IPsec)." .

sec:ComplianceStandard
  a rdfs:Class ;
  rdfs:label "Compliance Standard" ;
  rdfs:comment "Regulatory or industry standard (e.g., GDPR, HIPAA, ISO 27001)." .

% ==============================
% PROPERTIES
% ==============================

cloudeng:hasControlInterface
  a rdf:Property ;
  rdfs:domain cloudeng:CloudEngine ;
  rdfs:range cloudeng:ControlInterface ;
  rdfs:label "has control interface" .

cloudeng:hasBusinessInterface
  a rdf:Property ;
  rdfs:domain cloudeng:CloudEngine ;
  rdfs:range cloudeng:BusinessInterface ;
  rdfs:label "has business interface" .

cloudeng:hasAuditInterface
  a rdf:Property ;
  rdfs:domain cloudeng:CloudEngine ;
  rdfs:range cloudeng:AuditInterface ;
  rdfs:label "has audit interface" .

cloudeng:hasDataInterface
  a rdf:Property ;
  rdfs:domain cloudeng:CloudEngine ;
  rdfs:range cloudeng:DataInterface ;
  rdfs:label "has data interface" .

sec:hasSecurityPolicy
  a rdf:Property ;
  rdfs:domain cloudeng:CloudEngine ;
  rdfs:range sec:SecurityPolicy .

sec:usesIdentityProvider
  a rdf:Property ;
  rdfs:domain cloudeng:Interface ;
  rdfs:range sec:IdentityProvider .

sec:supportsAuthentication
  a rdf:Property ;
  rdfs:domain cloudeng:Interface ;
  rdfs:range sec:AuthenticationMechanism .

sec:enforcesAuthorization
  a rdf:Property ;
  rdfs:domain cloudeng:Interface ;
  rdfs:range sec:AuthorizationMechanism .

sec:encryptsData
  a rdf:Property ;
  rdfs:domain cloudeng:Interface ;
  rdfs:range sec:EncryptionMethod .

sec:encryptionScope
  a rdf:Property ;
  rdfs:domain sec:EncryptionMethod ;
  rdfs:range sec:EncryptionScope .

sec:usesTransportSecurity
  a rdf:Property ;
  rdfs:domain cloudeng:Interface ;
  rdfs:range sec:TransportSecurityProtocol .

sec:compliesWith
  a rdf:Property ;
  rdfs:domain sec:SecurityPolicy ;
  rdfs:range sec:ComplianceStandard .

sec:implementsStandard
  a rdf:Property ;
  rdfs:domain [ rdfs:subClassOf rdfs:Resource ] ;
  rdfs:range sec:ComplianceStandard ;
  rdfs:label "implements or satisfies a compliance standard" .

% ==============================
% SECURITY INSTANCES
% ==============================

% Identity Providers
sec:Keycloak
  a sec:IdentityProvider ;
  rdfs:label "Keycloak" .

sec:Okta
  a sec:IdentityProvider ;
  rdfs:label "Okta" .

% Authentication Mechanisms
sec:OAuth2
  a sec:AuthenticationMechanism ;
  rdfs:label "OAuth 2.0" ;
  rdfs:comment "Open authorization protocol for delegated access" ;
  sec:implementsStandard iso27001:A.9.2.2,    # User access provisioning
                         iso27001:A.9.4.2,    # Secure log-on procedures
                         csa:IVS-03,          # Password Management
                         csa:IVS-09,          # Strong Authenticators
                         nist80053:IA-2,      # Identification and Authentication
                         nist80053:IA-3 .     # Device Identification and Authentication

sec:SAML
  a sec:AuthenticationMechanism ;
  rdfs:label "SAML 2.0" ;
  rdfs:comment "Federated identity protocol for single sign-on and attribute assertions" ;
  sec:implementsStandard iso27001:A.9.2.2, iso27001:A.9.4.2, nist80053:IA-2 .

sec:APIKey
  a sec:AuthenticationMechanism ;
  rdfs:label "API Key" ;
  rdfs:comment "Shared secret or credential used by services and automation; should be rotated and scoped" ;
  sec:implementsStandard iso27001:A.9.2.3, nist80053:AC-2 .

sec:X509Cert
  a sec:AuthenticationMechanism ;
  rdfs:label "X.509 Certificate" ;
  rdfs:comment "Public key certificates for mutual TLS and service authentication" ;
  sec:implementsStandard iso27001:A.10.1.1, nist80053:IA-5 .

% Authorization Mechanisms
sec:RBAC
  a sec:AuthorizationMechanism ;
  rdfs:label "Role-Based Access Control" ;
  rdfs:comment "Coarse-grained access control by roles and role assignments; commonly used in OpenStack and cloud IAMs" ;
  sec:implementsStandard nist80053:AC-3, iso27001:A.9.4.1, csa:IVS-02 .

sec:ABAC
  a sec:AuthorizationMechanism ;
  rdfs:label "Attribute-Based Access Control" ;
  rdfs:comment "Policy decisions based on attributes of subjects, objects, and environment; useful for fine-grained controls" ;
  sec:implementsStandard nist80053:AC-4, iso27001:A.9.4.1 .

sec:OAuth2Scopes
  a sec:AuthorizationMechanism ;
  rdfs:label "OAuth 2.0 Scopes" ;
  rdfs:comment "Authorization scopes used to limit delegated access in OAuth flows" ;
  sec:implementsStandard iso27001:A.9.4.2, nist80053:AC-3 .

% Encryption & Transport
sec:AES256
  a sec:EncryptionMethod ;
  rdfs:label "AES-256" ;
  rdfs:comment "Symmetric encryption algorithm commonly used for data-at-rest" ;
  sec:encryptionScope sec:AtRest ;
  sec:implementsStandard nist80053:SC-13, iso27001:A.10.1.1, csa:DCS-07 .

sec:TLS13
  a sec:EncryptionMethod ;
  rdfs:label "TLS 1.3" ;
  rdfs:comment "Transport Layer Security for protecting data in transit; preferred modern protocol" ;
  sec:encryptionScope sec:InTransit ;
  sec:implementsStandard nist80053:SC-13, iso27001:A.10.1.1, gdpr:Article32 .

sec:AtRest
  a sec:EncryptionScope ;
  rdfs:label "At Rest" ;
  rdfs:comment "Encryption applied to stored data, including object, block, or database storage" .

sec:InTransit
  a sec:EncryptionScope ;
  rdfs:label "In Transit" ;
  rdfs:comment "Encryption applied to data while moving across networks or between services" .

sec:TLS
  a sec:TransportSecurityProtocol ;
  rdfs:label "TLS" ;
  rdfs:comment "Transport security protocol family" ;
  sec:implementsStandard nist80053:SC-13, gdpr:Article32, iso27001:A.10.1.1 .

sec:IPsec
  a sec:TransportSecurityProtocol ;
  rdfs:label "IPsec" ;
  rdfs:comment "Network-layer transport security for site-to-site or host-to-host tunnels" .

% ==============================
% INDUSTRY STANDARDS (as ComplianceStandard instances)
% ==============================

% ISO/IEC 27001:2022
iso27001:A.9.4.1
  a sec:ComplianceStandard ;
  rdfs:label "ISO/IEC 27001: A.9.4.1 - Information access restriction" .

iso27001:A.10.1.1
  a sec:ComplianceStandard ;
  rdfs:label "ISO/IEC 27001: A.10.1.1 - Cryptographic controls policy" .

iso27001:A.12.4.1
  a sec:ComplianceStandard ;
  rdfs:label "ISO/IEC 27001: A.12.4.1 - Event logging" .

% NIST SP 800-53 Rev. 5
nist80053:AC-3
  a sec:ComplianceStandard ;
  rdfs:label "NIST SP 800-53 AC-3 - Access Enforcement" .

nist80053:SC-13
  a sec:ComplianceStandard ;
  rdfs:label "NIST SP 800-53 SC-13 - Cryptographic Protection" .

nist80053:AU-2
  a sec:ComplianceStandard ;
  rdfs:label "NIST SP 800-53 AU-2 - Audit Events" .

% CSA CCM v4
csa:IVS-02
  a sec:ComplianceStandard ;
  rdfs:label "CSA CCM IVS-02 - Identity and Access Management" .

csa:DCS-07
  a sec:ComplianceStandard ;
  rdfs:label "CSA CCM DCS-07 - Data Security and Information Lifecycle Management" .

% AWS Well-Architected Framework
aws:SecurityPillar
  a sec:ComplianceStandard ;
  rdfs:label "AWS Well-Architected Framework: Security Pillar" .

aws:SEC02
  a sec:ComplianceStandard ;
  rdfs:label "AWS WAF SEC02 - Enable traceability" .

aws:SEC03
  a sec:ComplianceStandard ;
  rdfs:label "AWS WAF SEC03 - Apply security at all layers" .

% GDPR
gdpr:Article32
  a sec:ComplianceStandard ;
  rdfs:label "GDPR Article 32 - Security of processing" .

% ==============================
% CLOUD PROVIDER IMPLEMENTATIONS
% ==============================

% OpenStack
openstack:Keystone
  a sec:IdentityProvider, cloudeng:ControlInterface ;
  rdfs:label "OpenStack Keystone" ;
  sec:supportsAuthentication sec:OAuth2, sec:APIKey ;
  sec:enforcesAuthorization sec:RBAC ;
  sec:usesTransportSecurity sec:TLS ;
  sec:implementsStandard iso27001:A.9.4.1, nist80053:AC-3, csa:IVS-02 .

openstack:Ceilometer
  a cloudeng:AuditInterface ;
  rdfs:label "OpenStack Ceilometer" ;
  sec:usesTransportSecurity sec:TLS ;
  sec:implementsStandard iso27001:A.12.4.1, nist80053:AU-2 .

% AWS
aws:IAM
  a sec:IdentityProvider, cloudeng:BusinessInterface ;
  rdfs:label "AWS Identity and Access Management (IAM)" ;
  sec:supportsAuthentication sec:APIKey, sec:X509Cert ;
  sec:enforcesAuthorization sec:RBAC, sec:ABAC ;
  sec:usesTransportSecurity sec:TLS ;
  sec:implementsStandard aws:SEC03, csa:IVS-02, nist80053:AC-3 .

aws:CloudTrail
  a cloudeng:AuditInterface ;
  rdfs:label "AWS CloudTrail" ;
  sec:usesTransportSecurity sec:TLS ;
  sec:implementsStandard aws:SEC02, iso27001:A.12.4.1, nist80053:AU-2 .

aws:S3
  a cloudeng:DataInterface ;
  rdfs:label "Amazon S3" ;
  sec:supportsAuthentication sec:APIKey ;
  sec:enforcesAuthorization sec:RBAC ;
  sec:encryptsData sec:AES256 ;
  sec:usesTransportSecurity sec:TLS ;
  sec:implementsStandard aws:SEC03, csa:DCS-07, iso27001:A.10.1.1, nist80053:SC-13 .

% ==============================
% GENERIC INTERFACE EXAMPLES (with security)
% ==============================

cloudeng:OCCI
  a cloudeng:ControlInterface ;
  rdfs:label "OCCI" ;
  sec:usesIdentityProvider sec:Keycloak ;
  sec:supportsAuthentication sec:OAuth2, sec:APIKey ;
  sec:enforcesAuthorization sec:RBAC ;
  sec:usesTransportSecurity sec:TLS ;
  sec:encryptsData sec:TLS13 .

cloudeng:SSOService
  a cloudeng:BusinessInterface ;
  sec:usesIdentityProvider sec:Okta ;
  sec:supportsAuthentication sec:SAML, sec:OAuth2 ;
  sec:enforcesAuthorization sec:OAuth2Scopes ;
  sec:usesTransportSecurity sec:TLS .

cloudeng:Syslog
  a cloudeng:AuditInterface ;
  rdfs:comment "Assumes syslog over TLS (RFC 5425)" ;
  sec:usesTransportSecurity sec:TLS ;
  sec:encryptsData sec:TLS13 ;
  sec:implementsStandard nist80053:AU-2 .

cloudeng:Swift
  a cloudeng:DataInterface ;
  rdfs:label "OpenStack Swift" ;
  sec:encryptsData sec:AES256 ;
  sec:usesTransportSecurity sec:TLS ;
  sec:implementsStandard csa:DCS-07, iso27001:A.10.1.1 .

% ==============================
% SECURITY POLICIES
% ==============================

sec:EnterpriseCloudPolicy
  a sec:SecurityPolicy ;
  sec:compliesWith 
    iso27001:A.9.4.1,
    iso27001:A.10.1.1,
    iso27001:A.12.4.1,
    nist80053:AC-3,
    nist80053:SC-13,
    nist80053:AU-2,
    csa:IVS-02,
    csa:DCS-07,
    gdpr:Article32,
    aws:SecurityPillar ;
  rdfs:comment "Comprehensive policy aligned with major cloud and security standards." .

% ==============================
% CLOUD ENGINE INSTANCES
% ==============================

cloudeng:SecureCloudEngine
  a cloudeng:CloudEngine ;
  cloudeng:hasControlInterface cloudeng:OCCI ;
  cloudeng:hasBusinessInterface cloudeng:SSOService ;
  cloudeng:hasAuditInterface cloudeng:Syslog ;
  cloudeng:hasDataInterface cloudeng:S3, cloudeng:Swift ;
  sec:hasSecurityPolicy sec:EnterpriseCloudPolicy .

cloudeng:HybridCompliantEngine
  a cloudeng:CloudEngine ;
  cloudeng:hasControlInterface openstack:Keystone ;
  cloudeng:hasBusinessInterface aws:IAM ;
  cloudeng:hasAuditInterface aws:CloudTrail ;
  cloudeng:hasDataInterface aws:S3 ;
  sec:hasSecurityPolicy sec:EnterpriseCloudPolicy ;
  rdfs:comment "Hybrid cloud engine compliant with ISO, NIST, CSA, GDPR, and AWS best practices." .
\end{lstlisting}

\section{Validation examples and OpenStack extraction}
This appendix contains a few practical validation examples (SPARQL and SHACL) and short commands to extract facts from an OpenStack deployment for instantiation.

\subsection{SPARQL: Check DataInterfaces declare encryption}
\begin{verbatim}
PREFIX cloudeng: <http://example.org/cloudengine#>
PREFIX sec: <http://example.org/security#>

SELECT ?data
WHERE {
  ?data a cloudeng:DataInterface .
  FILTER NOT EXISTS { ?data sec:encryptsData ?enc }
}
\end{verbatim}

\subsection{SHACL: Require encryption for DataInterface}
\begin{verbatim}
@prefix sh: <http://www.w3.org/ns/shacl#> .
@prefix cloudeng: <http://example.org/cloudengine#> .
@prefix sec: <http://example.org/security#> .

cloudeng:DataInterfaceShape
  a sh:NodeShape ;
  sh:targetClass cloudeng:DataInterface ;
  sh:property [
    sh:path sec:encryptsData ;
    sh:minCount 1 ;
    sh:message "Data interfaces must declare an encryption method (at-rest)." ;
  ] .
\end{verbatim}

\subsection{OpenStack extraction (examples)}
Use these commands on a machine with the OpenStack client configured (an active RC file / environment):

\begin{verbatim}
# List endpoints and save JSON
openstack endpoint list -f json > endpoints.json

# List projects, users, and role assignments
openstack project list -f json > projects.json
openstack user list -f json > users.json
openstack role assignment list --role <role-name> -f json > assignments.json

# Get service versions (example: Keystone)
openstack --os-identity-api-version 3 token issue

# Dump Swift account/container policies or metadata via swift CLI
swift stat account > swift_account.txt
swift stat container_name > container_meta.txt

# Example: dump policy file from a controller node (path may vary)
sudo cat /etc/nova/policy.json > nova_policy.json
\end{verbatim}

After collecting JSON/text artifacts, transform them into RDF triples. A minimal Python script using rdflib can perform this mapping; the script should produce Turtle that includes service endpoints, versions, role assignments, policy file hashes, and KMS references.

\bibliographystyle{plain}
\bibliography{references}

\end{document}